\documentclass{cupconf}
\usepackage{natbib}
\usepackage{aas_macros}
\usepackage{graphicx}

\let\realverbatim=\verbatim
\let\realendverbatim=\endverbatim
\renewcommand\verbatim{\par\addvspace{6pt plus 2pt minus 1pt}\realverbatim}
\renewcommand\endverbatim{\realendverbatim\addvspace{6pt plus 2pt minus 1pt}}
\makeatletter
\newcommand\verbsize{\@setfontsize\verbsize{10}\@xiipt}
\renewcommand\verbatim@font{\verbsize\normalfont\ttfamily}
\makeatother

  \checkfont{eurm10}
  \iffontfound
    \IfFileExists{upmath.sty}
      {\typeout{^^JFound AMS Euler Roman fonts on the system,
                   using the 'upmath' package.^^J}%
       \usepackage{upmath}}
      {\typeout{^^JFound AMS Euler Roman fonts on the system, but you
                   don't seem to have the}%
       \typeout{'upmath' package installed. CUPconf.cls can take advantage
                 of these fonts,^^Jif you use 'upmath' package.^^J}%
      }
  \else
  \fi

  \checkfont{msam10}
  \iffontfound
    \IfFileExists{amssymb.sty}
      {\typeout{^^JFound AMS Symbol fonts on the system, using the
                'amssymb' package.^^J}%
       \usepackage{amssymb}%

      }{}
  \fi

  \IfFileExists{amsbsy.sty}
    {\typeout{^^JFound the 'amsbsy' package on the system, using it.^^J}%
     \usepackage{amsbsy}}
    {}

\newsavebox{\astrutbox}
\sbox{\astrutbox}{\rule[-5pt]{0pt}{20pt}}

\def\spose#1{\hbox to 0pt{#1\hss}}
\def\lta{\mathrel{\spose{\lower 3pt\hbox{$\mathchar"218$}}
           \raise 2.0pt\hbox{$\mathchar"13C$}}}
\def\gta{\mathrel{\spose{\lower 3pt\hbox{$\mathchar"218$}}
           \raise 2.0pt\hbox{$\mathchar"13E$}}}

\newcommand{\msun}{\,{\rm M_\odot}}

  \title[Evolution of massive black holes]{Evolution of massive black holes}

  \author[M. Volonteri]{Marta Volonteri $^1$}

  \affiliation{$^1$University of Michigan, Ann Arbor, MI 48109, USA}

\begin{document}

\maketitle

\begin{abstract}
Supermassive black holes are nowadays believed to reside in most local galaxies.  Accretion of gas and black hole mergers play a fundamental role in determining the two parameters defining a black hole: mass and spin.  I briefly review here some of the physical processes that are conducive to the evolution of the massive black hole population. I'll discuss black hole formation processes that are likely to place at early cosmic epochs, and how massive black hole evolve in a hierarchical Universe. The mass of the black holes that we detect today in nearby galaxy has mostly been accumulated by accretion of gas. While black hole--black hole mergers do not contribute substantially to the final mass of massive black holes, they influence the occupancy of galaxy centers by black hole, owing to the chance of merging black holes being kicked from their dwellings due to the ``gravitational recoil". Similarly, accretion leaves a deeper imprint on the distribution of black hole spins than black hole mergers do. The differences in accretion histories for black holes hosted in elliptical or disc galaxies may reflect on different spin distributions.

\end{abstract}
\firstsection 
\section{Introduction}
Black holes (BHs), as physical entities, span the full range of masses, from tiny BHs predicted by string theory, to monsters weighting by themselves almost as much as a dwarf galaxy (massive black holes, MBHs). Notwithstanding the several orders of magnitude difference between the smallest and the largest BH known, all of them can be described by only three parameters: mass, spin and charge. Astrophysical BHs are even simpler system, as charge can be neglected as well. The interaction between astrophysical black holes and their environment is where complexity enters the game. I will focus here on the formation and evolution of MBHs, with masses above thousands solar masses, and how we believe their evolution is symbiotic with that of their host.  

Let's start by recalling that MBHs in galaxy centers are far from being really ``black": we can easily trace the presence, as they are the engines powering the luminous quasars that have been detected up to high redshift. Nowadays we can detect in neighboring galaxies the dead remnants of this bright past activity. It is indeed well established that the centers of most local galaxies host MBHs with masses in the range $M_{BH} \sim 10^6-10^9\,M_\odot$ \citep[e.g.,][]{Ferrarese2000,Kormendy2001,Richstone1998}. The MBH population may extend down to the smallest masses.  Observationally, the record for the smallest MBH mass belongs to the dwarf Seyfert~1 galaxy POX 52 is thought to contain a BH of mass $M_{BH} \sim 10^5\,M_\odot$ \citep{Barth2004}. At the other end, however, the Sloan Digital Sky survey detected luminous quasars at very high redshift, $z>6$. Follow-up observations confirmed that at least some of these quasars are powered by supermassive black holes with masses $\simeq 10^9\, M_\odot$ \citep{Barthetal2003,Willottetal2005}. We are therefore left with the task of explaining the presence of very big MBHs when the Universe is less than {\rm 1 Gyr} old, and of much smaller BHs lurking in {\rm 13 Gyr} old galaxies. 

\section{Massive black holes in a hierarchical Universe} 
The demography of massive black holes in the local Universe has been clarified in the last ten years by studies of the central regions of relatively nearby galaxies (mainly with quiescent nuclei).  The mass of MBHs detected in neighboring galaxies scales with the bulge luminosity - or stellar velocity dispersion - of their host galaxy \citep{Ferrarese2000,Gebhardt2000,Tremaine2002}, suggesting a single mechanism for assembling black holes and forming spheroids in galaxy halos. The evidence is therefore in favour of a co-evolution between galaxies, black holes and quasars. In the currently favoured cold dark matter cosmogonies \citep{Spergel2007}, present-day galaxies have been assembled via a series of mergers, from small-mass building blocks which form at early cosmic times.  In this paradigm galaxies experience multiple mergers during their lifetime. If most  galaxies host BHs in their centre, and a local galaxy has been made up by multiple mergers, then a black hole binary is a natural evolutionary stage. During each galaxy merger event, the central black holes already present in each galaxy would be dragged to the centre of the newly formed galaxy  via dynamical friction ($\approx 0.1-10$ pc), and then if/when they get close ($\approx 0.01-0.001$ pc) the black hole binary would coalesce  via emission of gravitational radiation. The gap between where dynamical friction ceases to be efficient ($a_h\simeq Gm_2/(4\sigma_*^2)$) and where emission of gravitational waves takes over, at binary separations about two orders of magnitude smaller, could be the bottleneck of the merger process.

\begin{figure}
\includegraphics[width=\columnwidth]{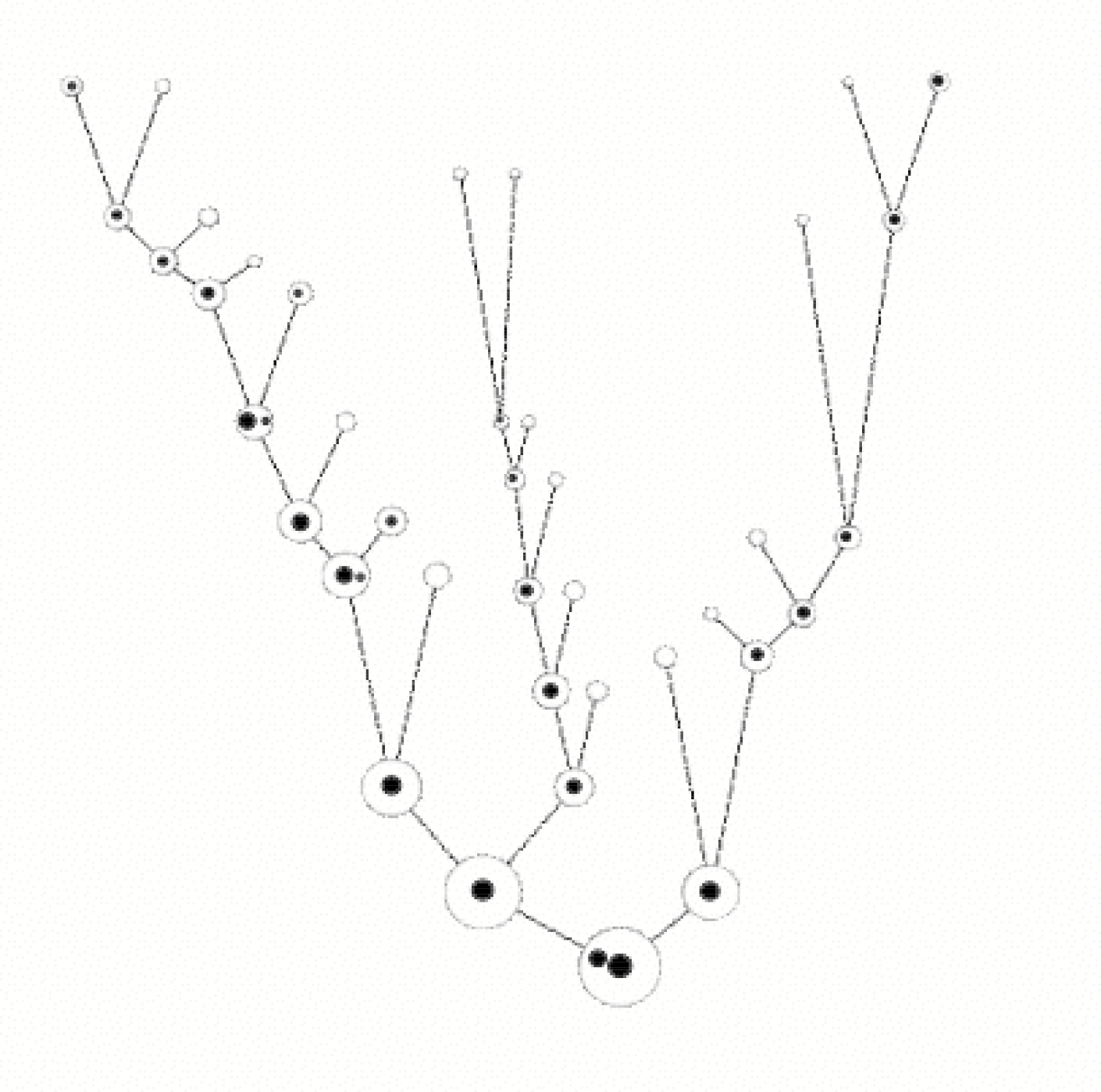}
\caption{A cartoon of the assembly of a galaxy and
its central black hole in cold-dark matter cosmology. 
Time increases from top to bottom, and every junction between two ``branches" of the merger tree marks a galaxy merger. The central MBHs are indicated by black dots (not in scale). In this case the final galaxy is assembled from the merger of twenty smaller galaxies, containing a total of four seed black holes, resulting in four mergers of binary black holes.}
\label{fig0}
\end{figure}

In gas-poor systems, the sub-parsec evolution of the binary while gravitational radiation emission is still negligible, may be largely determined by three-body interactions with background stars \citep{BBR1980}, by capturing the stars that pass within a distance of the order of the binary semi-major axis and ejecting them at much higher velocities \citep{Quinlan1996,Milosavljevic2001,Sesana2007}. Dark matter particles will be ejected by decaying binaries in the same way as the stars, i.e. through the gravitational slingshot. The hardening of the binary modifies the density profile, removing mass interior to the binary orbit, depleting the galaxy core of stars and dark matter, and slowing down further decay. We can use a toy model to understand the typical timescales \citep{VHM}. If we assume that the stellar mass removal scours a core of radius $r_c$ and constant 
density $\rho_c\equiv \rho_*(r_c)$ into a pre-existing isothermal sphere, the total mass ejected as the binary shrinks from $a_h$ to $a<a_h$ can be written as
\begin{equation}
{\cal M}_{\rm ej}= {4\over 3}{\sigma_*^2 (r_c)\over G}, 
\label{rcore}
\end{equation}
where $m_2$ is the least massive BH in the binary, and $\sigma_*$ is the velocity dispersion of the stellar system. 
The core radius then grows as 
\begin{equation}
r_c(t)\approx{3\over4 \sigma_*^2}G(m_1+m_2)\int_{a(t)}^{a_h}{\frac{1}{a}\,da}.\label{rc}
\end{equation}
The binary separation quickly falls below $r_c$ and subsequent evolution 
is slowed down due to the declining stellar density, with a hardening time
$t_h=|a/\dot a|=2\pi r_c(t)^2/(H\sigma_*a)$ that becomes increasingly long as the 
binary shrinks and $r_c(t)$ increases.

In gas rich systems, however, the orbital evolution of the central MBH is likely dominated by dynamical friction against the surrounding gaseous medium.  The available simulations  \citep{Escalaetal2004,Dottietal2006,Mayeretal2006} show that the binary can shrink to about parsec or slightly subparsec scale by dynamical friction against the gas, depending on the gas thermodynamics. The interaction between a BH binary and an accretion disc can also lead to a very efficient transport of angular momentum, and drive the secondary BH to the regime where emission of gravitational radiation dominates on short timescales, comparable to the viscous timescale \citep{ArmitageNarajan2005,GouldRix2000}.

The viscous timescale depends on the properties of the accretion disc and of the binary:
\begin{equation}
t_{\rm vis}=0.1\,{\rm Gyr}\,a_{\rm pc}^{3/2}\left(\frac{H}{R}\right)^{-2}_{0.1}\alpha^{-1}_{0.1}\left(\frac{m_1}{10^4 M_\odot}\right)^{-1/2},
\end{equation}
where $a_{\rm pc}$ is the initial separation of the binary when the secondary MBH starts interacting with the accretion disc in units of parsec, $(h/r)$ is the aspect ratio of the accretion disc, $h/r=0.1$ above, $\alpha$ is the Shakura \& Sunyaev viscosity parameter, $\alpha=0.1$ above, and $m_1$ is the mass of the primary MBH, in solar masses.
The emission of gravitational waves takes over the viscous timescales at a separation \citep{ArmitageNarajan2005}:
\begin{equation}
a_{\rm GW}=10^{-8}\, \rm {pc} \left(\frac{H}{R}\right)^{-16/5}_{0.1}\alpha_{0.1}^{-8/5}q_{0.1}^{3/5}\left(\frac{m_1}{10^4 M_\odot}\right),
\end{equation}
where  $q=m_2/m_1\lta1$,  is the binary mass-ratio. The timescale for coalescence by emission of gravitational waves from $a_{\rm GW}$ is much shorter than the Hubble time:
\begin{equation}
t_{\rm gr}=2.3\times 10^6 {\rm yr}\left(\frac{a(t)}{0.1{\rm pc}}\right)^4\,\left(\frac{m_1m_2(m1+m2)}{2\times10^{18}\msun}\right).
\end{equation}

Hence, the physical processes driving the evolution of MBH binaries are likely {\it redshift and environmental} dependent. Fast mergers are probably common in high redshift young galaxies, and sluggish binaries might reach an impasse in low-redshift gas poor spheroids.

\begin{figure}
\includegraphics[width=\columnwidth]{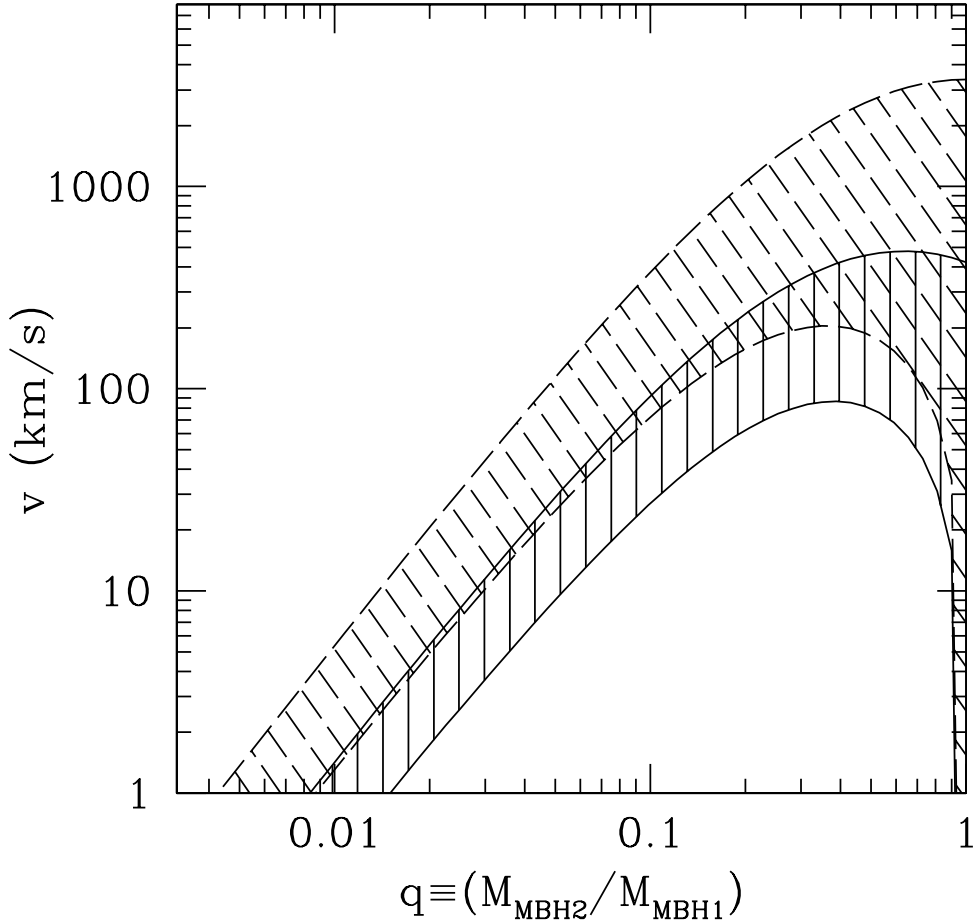}
\caption{Recoil velocity of spinning black holes as a function of binary mass ratio, $q$. {\it Solid curves:} spin axis aligned (or anti-aligned) with the orbital angular momentum (Baker et al. 2007). {\it Dashed curves:} spin axis in the orbital plane (Campanelli et al. 2007).  For every mass ratio we plot the combination of spins, $\hat a_1$ and $\hat a_2$ which minimizes ({\it lower curves}) or maximizes ({\it upper curves}) the recoil velocity. From Volonteri (2007)}
\label{fig1}
\end{figure}

Somehow counter-intuitively fast MBH mergers at very high redshift can bring an overall damage to the growth of the MBH population, rather than contribute to the build-up of more massive holes. This is due to the so-called ``gravitational recoil" (a.k.a. rocket, kick). When the members of a black hole binary coalesce, the center of mass of the coalescing system recoils due to the non-zero net linear momentum carried away by gravitational waves in the coalescence.  This recoil could be so violent that the merged hole breaks loose from shallow potential wells, especially in small mass pregalactic building blocks (Figure \ref{fig1}).
  
\begin{figure}
\includegraphics[width=\columnwidth]{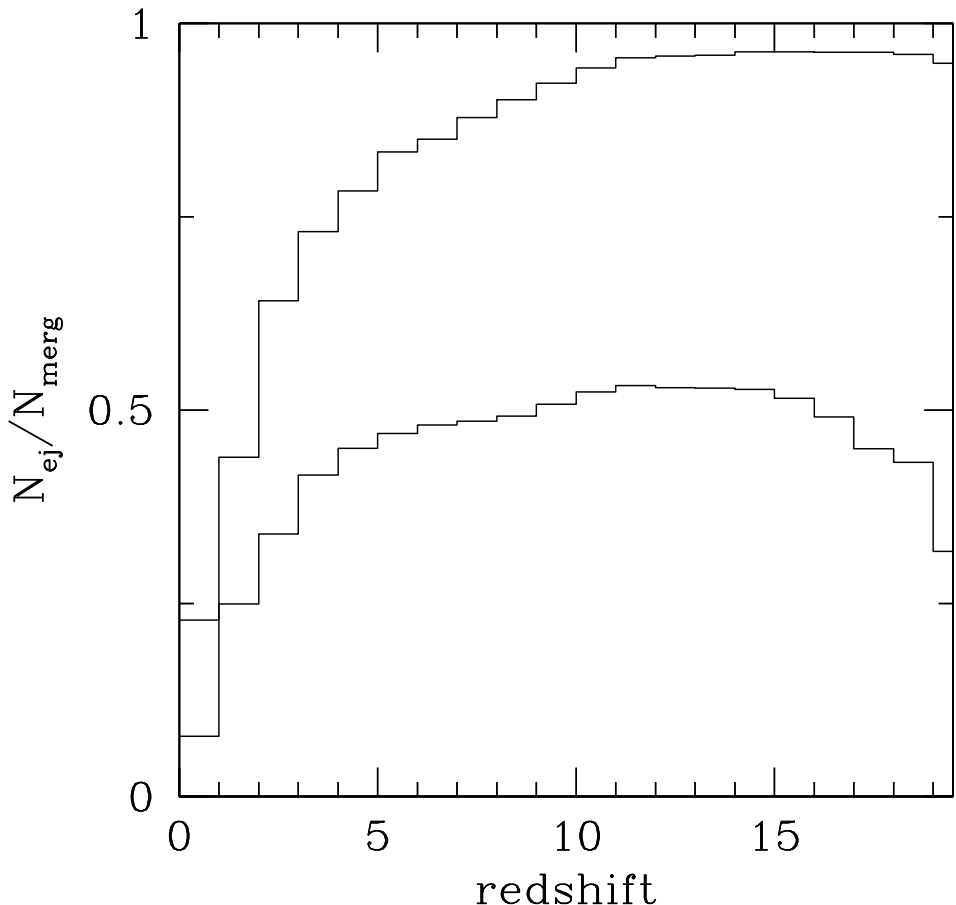}
\caption{\footnotesize Fraction of ejected binaries as a function of redshift for Schwarzschild MBHs ({\it lower histogram}) and spinning MBHs ({\it upper histogram}), assuming isotropic orbits. A binary is defined ejected if the recoil velocity is larger than the escape velocity from the host. }
\label{fig02}
\end{figure}

Comparing the recoil velocity to the escape velocity from their hosts, \cite{Volonteri2007} find that the fraction of ``lost'' binaries is very high ($>50-90\%$) at $z>10$, but it decreases at later times due to a combination of (i) the mass ratio distribution becoming shallower, and, (ii) the hierarchical growth of the hosts (Fig. ~\ref{fig02}). \cite{Schnittman2007} shows in a very elegant way that, even for large recoils, the very hierarchical nature of structure evolution ensures that a substantial fraction of galaxies retain their BHs, if evolution proceeds over a long series of mergers \citep[see also][]{menou2001}.  As, especially at high redshift, binaries represent the exception, rather than the rule, the possible ejection of most binaries before $z\simeq 5$ is not a threat to the evolution of the MBH population that has been detected in nearby galaxies.

Although the gravitational recoil does not damage to the evolution of the MBH population that we observe locally, it can be dangerous in very special cases. \cite{Haiman2004} pointed out that the recoil can be indeed threatening the growth of the MBHs that are believed to be powering the luminous quasars at $z\simeq 6$ detected in the Sloan survey \citep[e.g.,][]{Fanetal2001a}.  In fact, in such a biased volume, the density of halos where MBH formation can be efficient (either by direct collapse, or via PopIII stars) is highly enhanced.  
The net result is an higher fraction of binary systems, and binarity is especially common for the central galaxy of the main halo. While the ``average'' MBH experiences  at most one merger in its lifetime, a MBH hosted in a rare exceptionally massive halo can experience up to a few tens mergers, and the probability of ejecting the central MBH, halting its growth, is 50-80\% at $z>6$ \citep{VolonteriRees2006}. This implies that MBHs at high redshift did not mainly grow via mergers. 
  
\section{Scenarios for massive black hole formation}
A single big galaxy can be traced back to the stage when it was split up in hundreds of smaller components with individual internal velocity dispersions as low as 20 ${\rm km \,s^{-1}}$. Did black holes form with the same efficiency in small galaxies (with shallow potential wells), or did their formation had to await the buildup of substantial galaxies with deeper potential wells? 

The formation of massive black holes is far less understood than that of their light, stellar mass, counterparts. The "flowing chart" presented by \cite{Rees1978} still stands as a guideline for the possible paths leading to formation of massive BH seeds in the center of galactic structures. One first possibility is the direct formation of a BH from a collapsing gas cloud 
\citep{haehnelt1993,LoebRasio1994,Eisenstein1995,BrommLoeb2003,Koushiappas2004,BegelmanVolonteriRees2006,LN2006}. In the most common situations, rotational support can halt the collapse before densities required for MBH formation are reached. Halos, and their baryonic cores, possess in fact angular momentum, $J$, believed to be acquired by tidal torques due to interactions with neighboring halos. This can be quantified through the so-called spin parameter, which represents the degree of rotational support available in a gravitational system: $\lambda_{\rm spin} \equiv J |E|^{1/2}/G M_h^{5/2}$, where $E$ and $M_h$ are the total energy and mass of the halo.  

Let $f_{\rm gas}$ be the gas fraction of a proto-galaxy mass, and $f_d$ the fraction of the gas which can cool; a mass $M=f_d\,f_{\rm gas}\,M_h$ would then settle into a rotationally supported disc \citep{MoMaoWhite1998,OhHaiman2002} with a scale radius $\simeq  \lambda_{\rm spin} r_{\rm vir}$, where $r_{\rm vir}$ is the virial radius of the proto--galaxy. Spin parameters found in numerical simulations are distributed log-normally in $\lambda_{\rm spin}$, with mean  $\bar \lambda_{\rm spin}=0.04$ and standard deviation $\sigma_\lambda=0.5$ \citep[e.g.,][] {Bullock2001, VandenBosch2002}. The tidally induced angular momentum would therefore be enough to provide centrifugal support at a distance $\simeq 20$ pc from the center, and halt collapse.  Additional mechanisms inducing transport of angular momentum are needed to further condense the gas. 

The loss of angular momentum can be driven either by (turbulent) viscosity or by global dynamical instabilities, such as the "bars-within-bars" mechanism \citep{Shlosman1989,BegelmanVolonteriRees2006}.  The gas can therefore condense to form a central massive object, either a supermassive star, which eventually becomes subject to post-Newtonian gravitational instability and forms a seed BH, or via a low-entropy star-like configuration where a small black hole forms in the core and grows by accreting the surrounding envelope (Begelman, Rossi \& Armitage 2007).  The masses of the seeds predicted by different models vary, but they are typically in the range $M_{BH} \sim 10^4-10^6\,M_\odot$. 

Alternatively, the seeds of MBHs can be associated with the remnants of the first generation of stars, formed out of zero metallicity gas. The first stars are believed to form at $z\gta 10$ in halos which represent high-$\sigma$ peaks of the primordial density field.  The main coolant, in absence of metals, is molecular hydrogen, which is a rather inefficient coolant.  The inefficient cooling might lead to a very top-heavy initial stellar mass function, and in particular to the production of very massive stars with masses $>100 M_\odot$ (Carr, Bond, \& Arnett 1984). If very massive stars form above 260 $M_\odot$,  they would rapidly collapse to massive BHs
with little mass loss \citep{Fryer2001}, i.e., leaving behind seed BHs with masses $M_{BH} \sim 10^2-10^3\,M_\odot$ \citep{MadauRees2001,VHM}.  
  \begin{figure}
\includegraphics[width=\columnwidth]{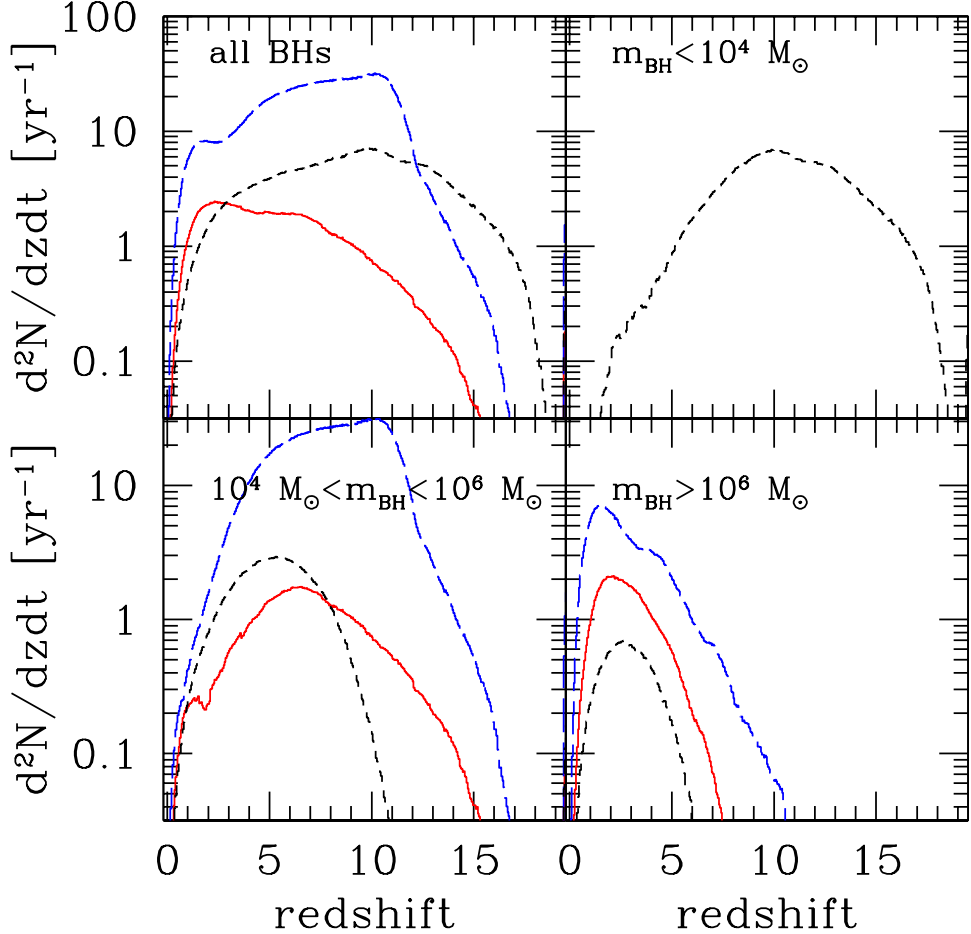}
  \caption{Predicted rate of MBH binary coalescences per unit redshift, in different MBH mass intervals. {\it Solid curve:} MBH seeds from PopIII stars. {\it Long dashed curve:} MBH seeds from direct collapse (Koushiappas et al. 2004).  {\it Short dashed curve:} MBH seeds from direct collapse (Begelman, Volonteri \& Rees 2006). Adapted from Sesana et al. 2007.}
\label{fig2}  
\end{figure}

\subsection{Observational tests of MBH formation scenarios}
What are the possible observational tests of MBH formation scenarios?
Detection of gravitational waves from seeds merging at the redshift of formation \citep{GW3} is probably one of the best ways to discriminate among formation mechanisms. The planned {\it Laser Interferometer Space Antenna} ({\it LISA}) in principle is sensible to gravitational waves from binary MBHs with masses in the range $10^3-10^6\; M_\odot$ basically at any redshift of interest.   A large fraction of coalescences will be directly observable by {\it LISA}, and on the basis of the detection rate, constraints can be put on the MBH formation process. Different theoretical models for the formation of MBH seeds and dynamical evolution of the binaries predict merger rates that largely vary one from the other (Figure ~\ref{fig2}, \citet{GW3}).   

\begin{figure}   
   \includegraphics[width=\columnwidth]{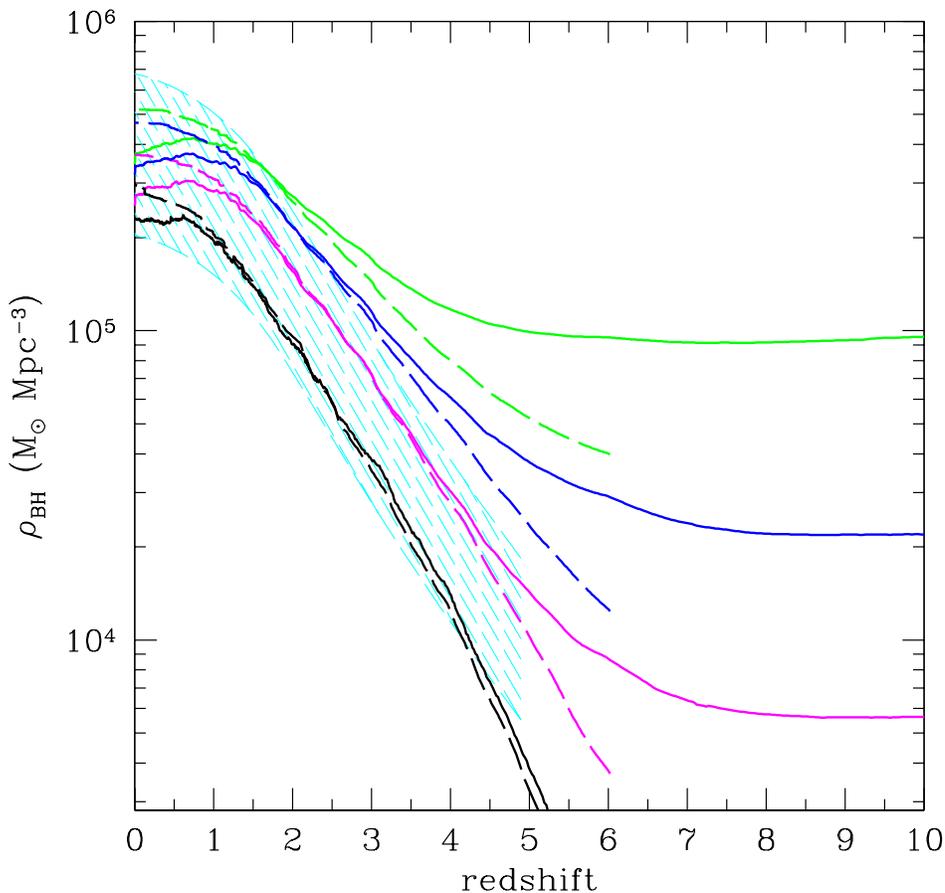} 
   \caption{Integrated black hole mass density as a function of
     redshift. Solid lines: total mass density locked into nuclear
     black holes.  Dashed lines: integrated mass density {\rm
     accreted} by black holes.  Models based on BH remnants of
     Population III stars (lowest curve), models based on direct collapse \citep	 {LN2006}, with different efficiencies.  
     Shaded area: constraints from Soltan-type
     arguments, where we have varied the radiative efficiency from a
     lower limit of 6\% (applicable to Schwarzschild MBHs, upper
     envelope of the shaded area), to about 20\% (Wang et
     al. 2006).}
   \label{fig5}
\end{figure}

The imprint of different formation scenarios can also be sought in observations at lower redshifts (Volonteri, Lodato \& Natarajan 2007).  Since during the quasar epoch MBHs increase their mass by a large factor, signatures of the seed formation mechanisms are likely more evident at {\it earlier epochs}. Fig.~\ref{fig5} compares the integrated comoving mass density in MBHs to the expectations from Soltan-type arguments (F. Haardt, private communication), assuming that quasars are powered by radiatively efficient flows (for details, see \citealt{YuTremaine2002,Elvis2002,Marconi2004}). The curves differ only with respect to the MBH formation scenario. We either assume that seeds are Population III remnants (black curve), or that seeds are formed via direct collapse with different efficiencies \citep{LN2006}. While during and after the quasar
epoch the mass densities in our theoretical models differ by less than a
factor of 2, at $z>3$ the differences become more pronounced.  The comoving mass density, an integral constraint, is reasonably well determined out to $z = 3$ but is still poorly known at higher redshifts. The increasing area and depth of high redshift survey, especially in X--rays, will increase the strength of our constraints \citep{Salvaterra2007}.

\begin{figure}
\includegraphics[width=\columnwidth]{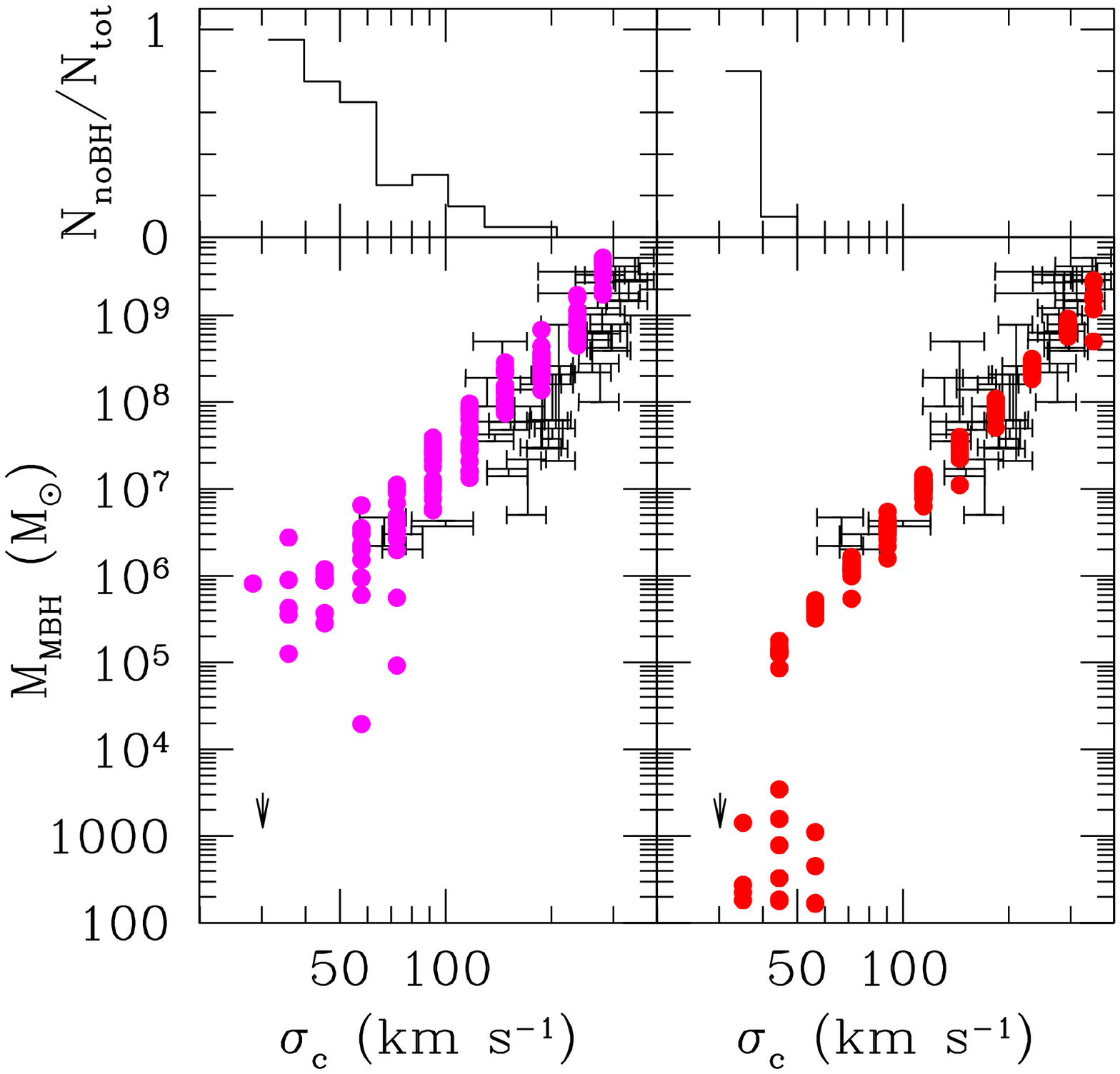}
\caption{The $M_{\rm bh}-$velocity dispersion ($\sigma_c$)
     relation at $z=0$. Every circle represents the central MBH in a
     halo of given $\sigma_c$.  Observational data are marked by their
     quoted errorbars, both in $\sigma_c$, and in $M_{\rm bh}$
     (Tremaine et al. 2002).  Left panel: direct collapse seeds, Population III star seeds.  {\it
     Top panels:} fraction of galaxies at a given velocity dispersion
     which {\bf do not} host a central MBH. Adapted from Volonteri, Lodato \& Natarajan (2007).}
\label{fig3}
\end{figure}

In our neighbourhood, the best diagnostic of MBH formation mechanisms would be the measure of MBH masses in low-luminosity galaxies. This can be understood in terms of the cosmological bias. The progenitors of massive galaxies (or clusters of galaxies) have both a high probability of hosting MBH seeds (cfr. \citealt{MadauRees2001}), and a high probability that the central MBH is not ``pristine", that is it has increased its mass by accretion, or it has experienced mergers and dynamical interactions.
In the case of low-bias systems, such as isolated dwarf galaxies, very
few of the high-$z$ progenitors have the deep potential wells needed
for gas retention and cooling, a prerequisite for MBH formation. The signature of the efficiency of the formation of MBH seeds will consequently be stronger in isolated dwarf galaxies. Hence, MBH formation models are distinguishable at the low mass end of the BH mass function, while at the high mass end the effect of initial seeds appears to be sub-dominant. The clearest signature of massive seeds, compared to Population III remnants, would be a lower limit of order the typical mass of seeds to the mass of MBHs in galaxy centers, as shown in Fig. ~\ref{fig3}.  Additionally, the fraction of galaxies without a MBH increases with decreasing halo masses at $z = 0$.  A larger fraction of low mass halos are devoid of central black holes for lower seed formation efficiencies. 
While current data in the low mass regime is still scant \citep{Barth2005}, future campaigns with the Giant Magellan Telescope or JWST are likely to probe this region of parameter space with significantly higher sensitivity.

\section{Accretion: mass growth and quasar activity} 
Accretion is inevitable during the ``active" phase of a galactic nucleus. Observations tell us that AGN are widespread in both the local and early Universe. All  the information that we have gathered on the evolution of MBHs is indeed due to studies of AGN, as we have to await for LISA to be able to ``observe" quiescent MBHs in the distant Universe. A key issue is then the relative importance of mergers and accretion in the build-up of the black holes, in dependence of the host properties (mass, redshift, environment). 

The accretion of mass at the Eddington rate would cause a black hole mass to increase in time as
\begin{equation}
M(t)=M(0)\,\exp\left(\frac{1-\epsilon}{\epsilon}\frac{t}{t_{\rm Edd}}\right),
\end{equation}
where $t_{\rm Edd}=0.45\,{\rm Gyr}$ and $\epsilon$ is the radiative efficiency. 
The classic argument of \cite{Soltan1982}, compares the total mass of black holes today with the total radiative output by known quasars,  by integration over redshift and luminosity of the luminosity function of quasars 
\citep{YuTremaine2002,Elvis2002,Marconi2004}.  The total energy density can then be converted  into the total mass density accreted by black holes during the active phase, by assuming a mass-to-energy conversion efficiency,  $\epsilon$ 
\citep{Aller2002,Merlonietal2004,Elvis2002,Marconi2004}.  The similarity of the total mass in MBHs today and the total mass accreted by MBHs  implies that the last 2-3 e-folds of the mass is grown via radiatively efficient accretion, rather than accumulated through mergers or radiatively inefficient accretion.  However,  most ot the Ôe-foldsÕ  (corresponding to a relatively small amount of mass, say the first  10\% of mass) could be gained rapidly via, e.g.,  radiatively inefficient accretion.  This argument is particularly important at early times.

\begin{figure}
\includegraphics[width=\columnwidth]{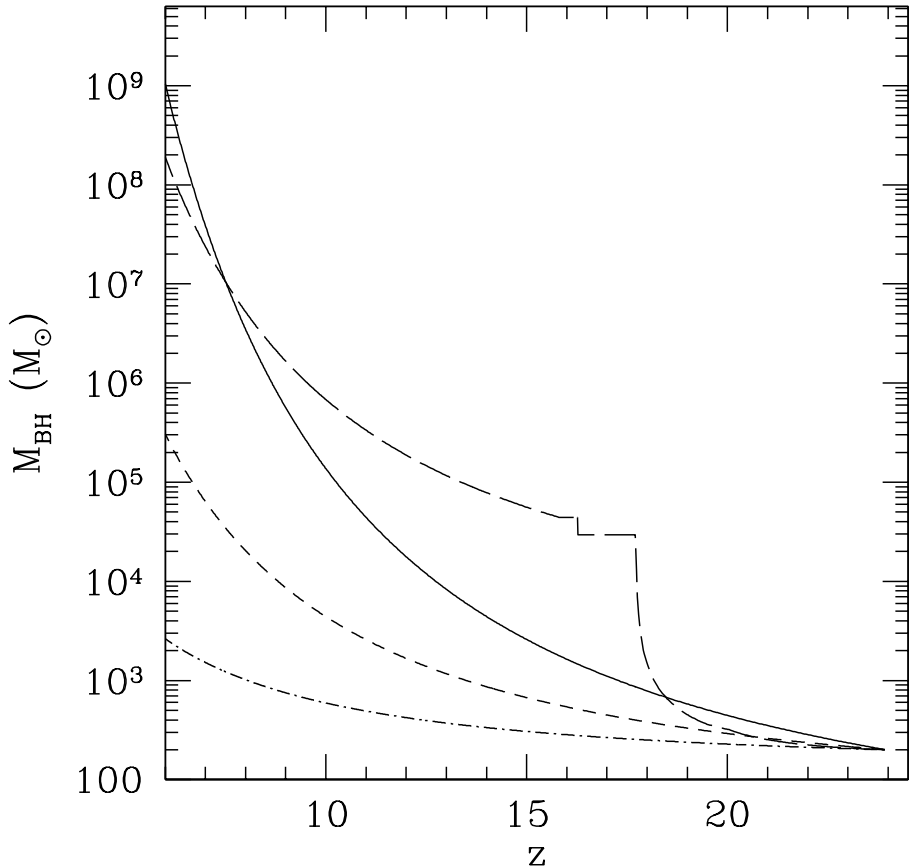}
  \caption{Growth of a MBH mass under different assumption for the accretion
rate and efficiency. Eddington limited accretion: $\epsilon=0.1$ ({\it solid line}),
$\epsilon=0.2$ ({\it short dashed line}), $\epsilon=0.4$ ({\it dot-dashed line}). Radiatively inefficient super-critical accretion, as in Volonteri \& Rees 2005 ({\it long dashed line}). }
\end{figure}

The Sloan Digital  Sky survey detected luminous quasars at very high redshift, $z>6$, when the Universe was less than {\rm 1 Gyr} old. Follow-up observations confirmed that at least some of  these quasars are powered by supermassive black holes with masses $\simeq 10^9\, M_\odot$ \citep{Barthetal2003,Willottetal2005}.  Given a seed mass $M(0)$ at $z=50$ or less, the higher the efficiency, the longer it takes for the MBH to grow in mass by (say) 10 e-foldings. If accretion is radiatively efficient, via a geometrically thin disc, the alignment of a MBH with the angular momentum of the accretion disc tends to efficiently spin holes up (see section 5), and radiative efficiencies can therefore approach 30-40\%. With such a high efficiency, $\epsilon=0.3$, it can take longer than  {\rm 2 Gyr} for the seeds to grow up to a billion solar masses.   

Let us consider the extremely rare high redshift (say $z>15$) metal--free halos with virial temperatures $T_{\rm vir} > 10^4$K where gas can cool even in the absence of ${\rm H_2}$ via neutral hydrogen atomic lines. The baryons can therefore collapse until angular momentum becomes important. 
Afterward, gas settles into a rotationally supported dense disc at the center of the halo \citep{MoMaoWhite1998,OhHaiman2002}. This gas can supply fuel for accretion onto a MBH within it.  
Estimating the mass accreted by the MBH within the Bondi-Hoyle formalism, the accretion rate is initially 
largely above the Eddington limit \citep{VolonteriRees2005}.  When the supply is super-critical the excess radiation can be trapped, as radiation pressure cannot prevent the accretion rate from being super-critical, while the emergent luminosity is still Eddington limited in case of spherical or quasi-spherical configurations \citep{Begelman1979,BegelmanMeier1982}.  In the spherical case, 
though this issue remains unclear, it still seems  possible that when the inflow rate is 
super-critical, the radiative efficiency drops so that the hole can accept the material without 
greatly exceeding the Eddington luminosity.  The  efficiency could be low either because most 
radiation is trapped and advected inward, or because the flow adjusts so that the material can 
plunge in from an orbit with small binding energy \citep{AbramowiczLasota1980}. 
The creation of a radiation-driven outflow, which can possibly stop the infall of material, is also a possibility.  If radiatively inefficient supercritical accretion requires metal-free conditions in exceedingly  rare massive halos, rapid early growth, therefore, can happen only for a tiny fraction 
of MBH seeds. These MBHs are those powering the most luminous high redshift quasars, and later on to be found in the most 
biased  halos today. The global MBH population, instead, evolves at a more quiet and slow pace.

\section{Probing the other hair of astrophysical black holes}
Astrophysical BHs are characterized by just 2 parameters, mass and spin, which is a measure of the angular momentum of the hole. The spin is typically expressed vis the dimensionless parameter $\hat a \equiv J_h/J_{max}=c \, J_h/G \, M_{\rm BH}^2$, where $J_h$ is the angular momentum of the black hole. 

The spin of a hole affects the efficiency of the ``classical" accretion processes themselves; the value of $\hat a$ in a Kerr BH also determines how much energy is in principle extractable from the hole itself.  Assuming that relativistic jets are powered by rotating black holes through the Blandford-Znajek mechanism,  the so-called ``spin paradigm" asserts that
powerful relativistic jets  are produced in AGN with fast rotating
black holes \citep{Blandford1990}. 

Spin-up is a natural consequence of prolonged disc-mode accretion: any hole that has (for instance) doubled its mass by capturing material with constant angular momentum axis would end up with spinning rapidly, close to the maximum allowed value \citep{Bardeen1970,Thorne1974}. However, when an accretion disc does not
lie in the equatorial plane of the BH, that is, when the angular
momentum of the accretion disc is misaligned with respect to the
direction of $J_h$, accretion of counter-rotating material can cause the spin-down of MBHs, at least under particular conditions. A misaligned disc is subject to the Lense-Thirring precession, which tends to align the inner parts of the disc with the the angular momentum of the black hole, causing the inclination angle between the angular momentum vectors to decrease with decreasing distance from the MBH, forcing the inner parts of the accretion disc to rotate in the equatorial plane of the MBH \citep{BardeenPetterson1975}. The orbits can be co-rotating or counter-rotating depending the value of $J_d/J_h$.

\cite{King2005} indeed argue that accretion of counter-rotating material  is more common than accretion of co-rotating material. The counter-alignment condition depends on the ratio $0.5\,J_d/J_h$, where $J_h$ and $J_d$ are the angular momenta of the hole and of the disc, to be compared with the cosine of the inclination angle, $\phi$.  If $\cos \phi<-0.5\,J_d/J_h$, the counter-alignment condition is satisfied. 

However, sustained accretion from a twisted disc would align the MBH spin (and the innermost equatorial disc) with the angular momentum vector of the disc at large radii \citep{Scheuer1996}. If the disc was initially counter-rotating with respect to the MBH, a complete overflip would eventually occur, and then accretion of co-rotating material would act to spin up the MBH \citep{Bardeen1970}. Early work by \cite{Moderski1998} concluded that the Bardeen-Petterson effect can be neglected because the alignment time \citep[$10^7$ years,][]{Rees1978} is longer than the duration of a single accretion event. Later, however, a series of papers revised the alignment timescale, suggesting that it could be much shorter \citep{Scheuer1996,NatarajanPringle1998}.  This framework was investigated by \cite{Volonterietal2005} who argue that the lifetime of quasars is long enough that angular momentum coupling
between black holes and accretion discs through the Bardeen-Petterson effect effectively forces the innermost region of accretion discs to align with black-hole spins (possibly through spin flips), and hence all AGN black-holes should have large spins \citep[see also][]{Cattaneo2002}.

In this context, \cite{VolonteriSikoraLasota2007} have explored the dependence of the alignment timescale in a  \cite{ShakuraSunyaev} disc on: viscosity $\nu_2$\footnote{The viscosity characterizing the alignment of the disc can be different from the accretion driving viscosity, $\nu_1$, which is responsible for the transfer of the component of the angular momentum parallel to the
spin of the disc. The relation between $\nu_1$ and $\nu_2$ is the
main uncertainty of the problem, assuming of course that such
two-viscosity description is adequate at all. Describing $\nu_1$ by
the Shakura--Sunyaev parameter $\alpha$ one can show \citep{PP1983} that the regime in which $H/R < \alpha \ll 1$ ($H$ being the disc thickness) one has $\nu_1/\nu_2\approx \alpha^2$.  However, for high accretion rates $\alpha \ll 1$ might not be appropriate, and in such a case $\nu_1$ is comparable to $\nu_2$ \citep{Kumar1985}.}, black hole mass $M_{\rm BH}$, Eddington ratio $f_{\rm Edd}$, accreted mass $\Delta m$. 

Within this simple picture, the timescale for disc-BH alignment can be estimated as 
\begin{equation}
t_{\rm align}\simeq\frac{J_h}{J_d(R_w)}t_{\rm acc}(R_w),
\label{eq:talign1}
\end{equation}
where $t_{\rm acc}$ is simply the accretion timescale, $t_{\rm acc}={R_w^2}/{\nu_1}$, and  $R_w$ marks the transition between (inner) alignment and (outer) misalignment. $R_w$ corresponds to the location in the disc where the timescale for radial diffusion of the warp is comparable to the local Lense-Thirring precession timescale \citep{Scheuer1996,NatarajanPringle1998}. The warp radius scales with the Schwarzschild radius of the BH as:
\begin{equation}
\frac{R_w}{R_s  }=
3.6\times 10^3 \hat
a^{5/8}\left(\frac{M_{\rm BH}}{10^8 \msun}\right)^{1/8}\,f_{\rm
Edd}^{-1/4}\left(\frac{\nu_2}{\nu_1}\right)^{-5/8}\alpha^{-1/2}.
\label{eq:rw}
\end{equation}

Defining the mass accreted during $t_{\rm align}$ as $m_{\rm
align}=t_{\rm align}\dot M$, one gets:
\begin{equation}
m_{\rm align}\simeq M_{\rm BH}\,\hat a \left(\frac{R_s}{R_w}\right)^{1/2}.
\label{eq:malign}
\end{equation}

Therefore, for all plausible assumptions, $m_{\rm align} \ll M_{\rm BH}$,
and a series of many randomly oriented accretion events with
accreted mass $\Delta m \ll m_{\rm align}$ should result in black-hole's spin oscillating
around zero. For the opposite case of $\Delta m \gg m_{\rm align}$ the
black hole will be spun-up to large positive spins; for $\Delta m \sim
M_{\rm BH}$ the hole will be spun-up to $\hat a \sim 1$.

\begin{figure}
\includegraphics[width=\columnwidth]{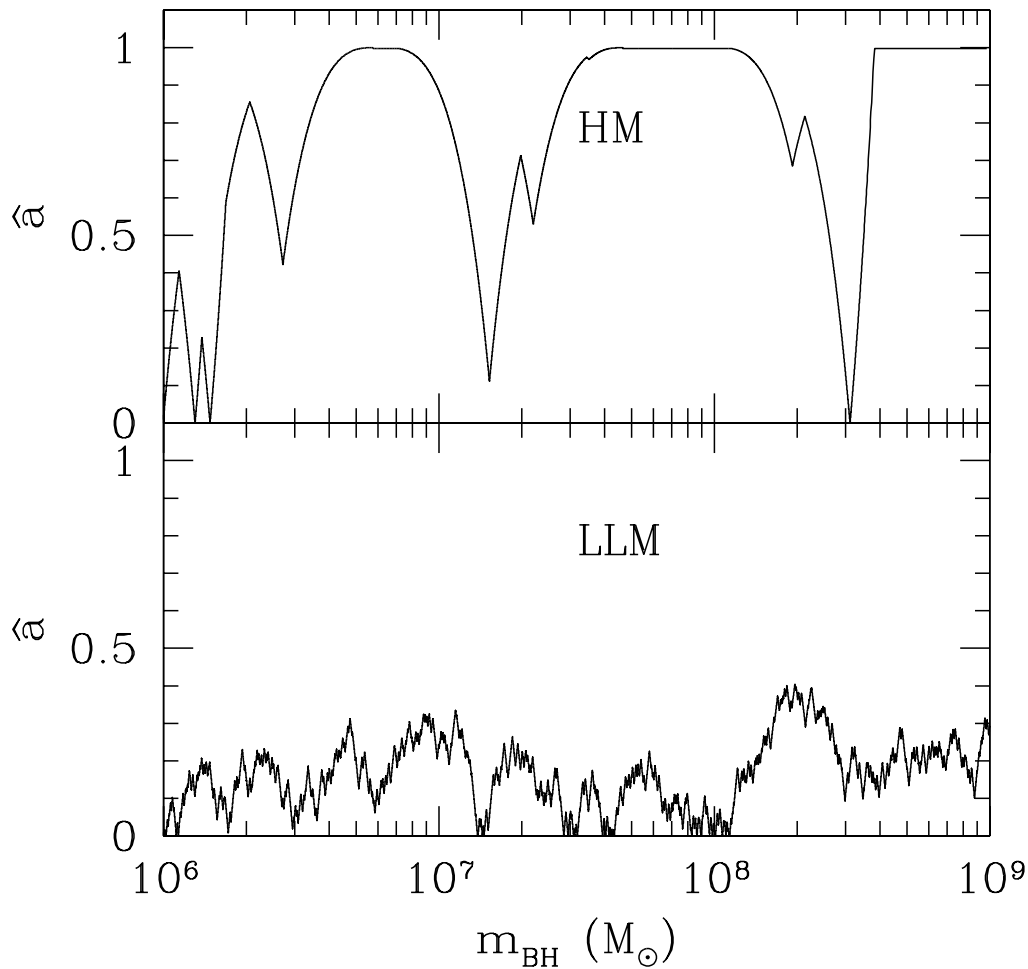}
\caption{Evolution of a MBH spin during a series of accretion episodes lasting for a total of a Hubble time. Initial mass $m_{\rm BH}=10^6 M_\odot$ , initial spin $\hat a=10^{-3}$. Lower panel:  the accreted mass at every accretion episode is constrained to be less than 0.01 of the MBH mass (LLM). Upper panel: the accreted mass is randomly extracted in the range 0.01-10 times the MBH mass (HM). Adapted from Volonteri, Sikora \& Lasota 2007.}
\label{fig4}
\end{figure}


However, both semi-analytical models of the cosmic MBH evolution
(Volonteri et al. 2005) and simulations of merger driven accretion
(di Matteo et al. 2005) show that most MBHs increase their mass by
an amount $\Delta m \gg m_{\rm align}$, if the evolution of the LF of quasars
is kept as a constraint. These high $\Delta m$ values are likely
characteristic of the most luminous quasars and most massive black
holes -- especially at high redshift. We expect therefore that
bright quasars at $z>3$ have large spins (upper panel in Fig. ~\ref{fig4}).  High spins in bright quasars are also indicated by the high radiative efficiency of quasars, as deduced from observations applying the Soltan argument \citep [and references therein ]{Soltan1982,Wang2006}.

\subsection{MBH spins and galaxy morphology}
If the events powering quasars coincide with the formation of elliptical galaxies (di Matteo et al. 2005), we might expect that
the MBH hosted by an elliptical galaxy had, as last major accretion
episode, a large increase in its mass. During this episode the spin increased significantly as well, possibly up to very high values. 

Black holes in spiral galaxies, on the other hand, probably had
their last major merger (i.e., last major accretion episode), if
any, at high redshift, so that enough time elapsed for the galaxy
disc to reform. Moreover, several observations suggest that single accretion
events last  $\simeq 10^5$ years in Seyfert galaxies, while the total activity lifetime (based on the fraction of disc galaxies that are Seyfert) is
$10^8-10^9$ years \citep[e.g.,][]{Kharb2006,Ho1997}. This
suggests that accretion events are very small and very
{`}compact'. Smaller MBHs, powering low luminosity Active Galactic Nuclei, likely grow by accreting smaller packets of material, such as
tidally disrupted stars \citep[for MBHs with mass $<2\times 10^6 \msun$,][]{
Milosavljevic2006}, or possibly molecular clouds \citep{HopkinsHernquist2006}.

Compact self-gravitating cores of molecular clouds (MC) can occasionally reach subparsec regions. Although the rate of such events is uncertain, we can adopt
the estimates of Kharb et al. (2006), and assume that about $10^4$ of
such events happen. We can further assume a lognormal distribution
for the mass function of MC close to galaxy centers \citep[based on the
Milky Way case, e.g.,][]{Perets2007}. We do not
distinguish here giant MC and clumps, and, for illustrative purpose
we assume a single lognormal distribution peaked at
$\log(M_{\rm MC}/\msun)=4$, with a dispersion of 0.75.

\begin{figure}
\includegraphics[width=\columnwidth]{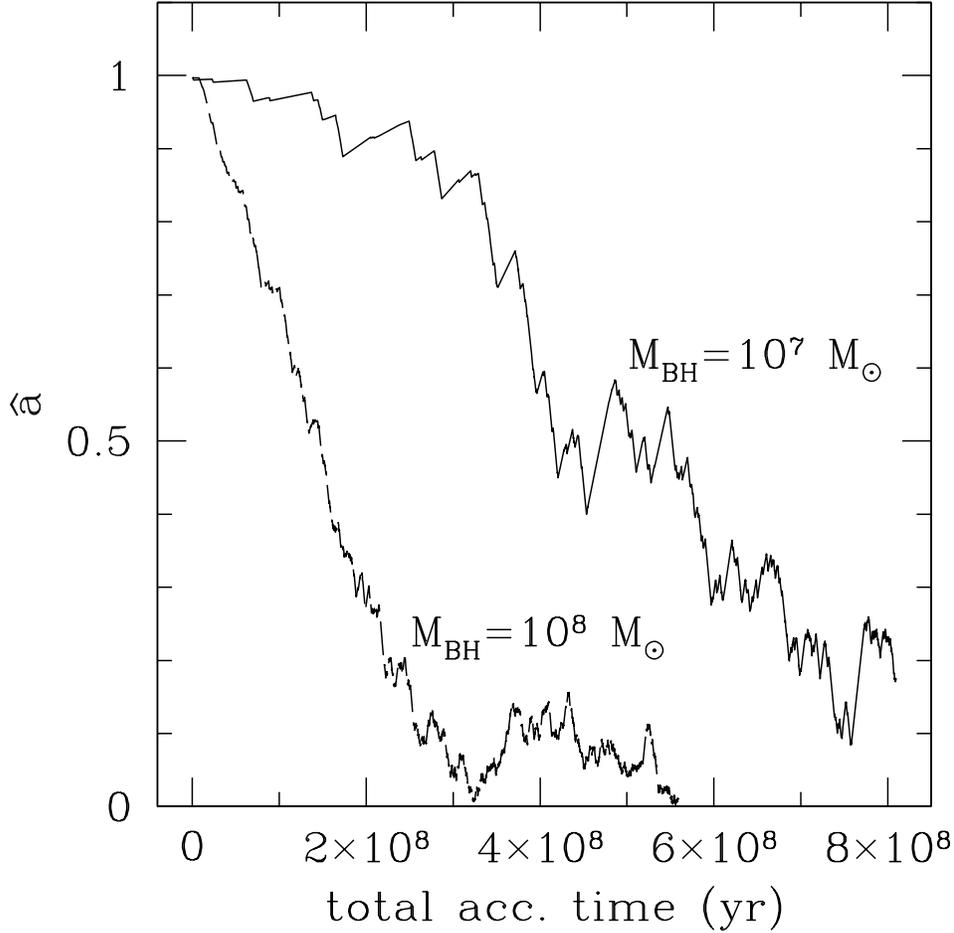}
\caption{Evolution of MBH spins due to accretion of molecular clouds
cores. We assume a lognormal distribution for the mass function of
molecular clouds peaked at $\log(M_{MC}/\msun)=4$, with a dispersion
of 0.75. The initial spin of the MBHs is 0.998. Upper curve: the
initial MBH mass is $10^7\msun$, lower curve: the initial MBH mass is
$10^8\msun$}
\label{fig6}
\end{figure}

Fig. \ref{fig6} shows the possible effect that accretion of molecular
clouds  can have on spinning MBHs. The result is, on the whole,
similar to that produced by minor mergers of black holes 
\citep{HB2003}, that is a spin down in a random walk fashion. 

In a gas-poor elliptical galaxy, however, substantial populations of molecular clouds are lacking (e.g. Sage et al. 2007), eliminating this channel of MBH feeding. Main sequence stars, however, linger in galaxy centers. Tidal disruption of stars is a feeding mechanism that has been proposed long ago \citep{Hills1975, Rees1988}. One expects discs formed by stellar debris to form with a random orientation. Stellar disruptions would therefore contribute to the spin-down of MBHs. The number of tidal
disruptions of solar type stars in an isothermal cusp per billion
years can be written as:

\begin{equation} N_*=4\times 10^5
\left(\frac{\sigma}{60\, {\rm km/s}}\right)\left(\frac{M_{\rm
BH}}{10^6 \msun}\right)^{-1}.
\end{equation}

Assuming that MBH masses scale with
the velocity dispersion, $\sigma$, of the galaxy (we adopt here the
Tremaine et al. 2002 scaling), we can derive the relative mass
increase for a MBH in 1 billion years:
\begin{equation} \frac{M_*}{M_{\rm
BH}}=0.37\left(\frac{M_{\rm BH}}{10^6\msun}\right)^{-9/8}.
\label{eqTD}
\end{equation}
The maximal level of spin down would occur assuming
that all the tidal disruption events form counterrotating discs,
leading to retrograde accretion. Eq. \ref{eqTD} shows that a small (say $10^6 \msun$) MBH starting at $\hat a=0.998$ would be spun down completely, on the other hand the spin of a larger (say $10^7 \msun$) MBH would not be changed
drastically. This feeding channel is likely efficient in early type discs which typically host faint bulges characterized by steep density cusps, both inside \citep{BahcallWolf1976} and outside \citep{Faber1997} the sphere of influence of the BH.  In this environment, the rate of stars which are tidally
disrupted by MBHs  \citep{Hills1975, Rees1988}  less massive than
$10^8\msun$\footnote{For black hole masses $\geqslant 2\times
10^8\msun$ the Schwarzschild radius exceeds the tidal disruption
radius for main-sequence stars.} is non negligible (eq. 5.5, Milosavljevic et
al. 2006). The situation is different for giant ellipticals: 
the central density profile often displays a shallow core, and
tidal disruption of stars is unlikely to play a dominant role.

Summarizing, the spin of MBHs in giant elliptical galaxies is likely dominated by massive accretion events which follow galaxy mergers. Both tidal disruption
of stars, and accretion of gaseous clouds is unlikely in shallow,
stellar dominated galaxy cores. The spin stays consequently high.  In a galaxy displaying instead power-law (cuspy) brightness profiles, the rate of stellar tidal disruptions is much higher and random small mass accretion events contribute to spin MBHs down. 

The different accretion histories in elliptical and disc galaxies seem to lead to a morphology~-~related bimodality of black-hole spin distribution
in the centers of galaxies: central black holes in giant elliptical
galaxies  may have (on average) much larger spins  than black holes in
spiral/disc galaxies. 

This result is in agreement with Sikora et al. (2007) who found that disc galaxies tend to be weaker radio sources with respect to elliptical hosts. \cite{Sikora2007} therefore proposed a revised ``spin paradigm", which incorporates elements of the ``accretion paradigm" \citep{Ulvestad2001, Merloni2003,Nipoti2005, Kording2006} according to which the radio-loudness is entirely related to the states of accretion discs, similarly to radio-emission from X-ray binaries \citep{Gallo2003}, related to transitions between two different accretion modes \citep{Livio2003}. It should be emphasized that even if the production of powerful relativistic jets is conditioned by the
presence of fast rotating BHs, it also depends on the accretion rate and on the presence of disc MHD winds required to provide the initial collimation of the central Poynting flux dominated outflow.

To assess the  role of spins in jet production, it is crucial to be able to measure MBH spins in AGN. An ingenious technique employs the Fe K$\alpha$ line at 6.4 keV in the X--ray spectrum  (2-10 keV), which is typically observed with broad asymmetric profile indicative of a relativistic disc \citep{Miller2007}. The iron line can in principle constrain the value of black hole spins \citep{Fabian1989,Laor1991}.  The value of $\hat a$  affects the location of the inner radius of the accretion disc (corresponding to the innermost stable circular orbit in the standard picture), which in turn has a large impact on the shape of the line profile,  because when the hole is rapidly rotating,  the emission is concentrated closer in,  and the line displays larger shifts. There is some evidence that this must be the case in some local AGN galaxies \citep{FabianMiniutti2005,Streblyanska2005,Comastri2006}, where the inferred location of the inner disc radius is well inside the maximum stable orbit for a Schwarzschild black hole, implying that these AGN should contain rotating Kerr MBH at their centers.  The assumption that the inner disc radius corresponds to the ISCO is not a trivial one, however, especially for thick discs \citep[see][]{Krolik1999,Afshordi2003}.  
 
Although observations of the iron line with the Chandra and XMM X-ray satellites are extending the studies of the innermost regions of MBHs, aiming at probing black hole properties, interpretation of these studies is impeded by the inherent 'messiness' of gas dynamics. A clean probe would be a compact mass in a precessing and decaying orbit around a massive hole. The detection of gravitational waves from a stellar mass BH, or even a white dwarf, or neutron star, falling into a massive black hole (Extreme Mass Ratio Inspiral, or EMRI)  can provide a unique tool to constrain the geometry of spacetime around BHs, and as a consequence, BH spins.   Indeed the spin is a measurable parameter, with a very high accuracy, in the gravitational waves {\it LISA} signal 
\citep{BarackCutler2004,Berti2005,Berti2006,Lang2006,Vecchio2004}. Gravitational waves from an EMRI can be used to map the spacetime of the central massive dark object. The resulting 'map' can tell us if the standard picture for the central massive object, a Kerr BH described by general relativity, holds.

\bibliography{../marta}
\bibliographystyle{cupconf}

\end{document}